\documentclass[a4paper,11pt]{article}
\pdfoutput=1 

\usepackage{jheppub} 

\usepackage[T1]{fontenc} 
\usepackage{txfonts}
\usepackage{braket}
\usepackage{latexsym}
\usepackage{wrapfig}
\usepackage{tabularx}
\usepackage{comment}
\newcommand{\der}{\partial}

\newcommand{\no}{\nonumber}

\newcommand{\im}{\mathrm{Im}}

\title{Extension of positivity bounds to non-local theories: IR obstructions to Lorentz invariant UV completions}

\author[a]{Junsei Tokuda}

\affiliation[a]{Department of Physics, Kyoto University,
\\Kyoto 606-8502, Japan}

\emailAdd{tokuda@tap.scphys.kyoto-u.ac.jp}

\abstract{We derive positivity bounds on low energy effective field theories which admit gapped, analytic, unitary, Lorentz invariant, and possibly {\it non-local} UV completions, by considering 2 to 2 scatterings of Jaffe fields whose Lehmann-K{\"{a}}ll{\'{e}}n spectral density can grow exponentially. Several properties of S-matrix, such as analyticity properties, are assumed in our derivation. 
Interestingly, we find that some of the positivity bounds obtained in the literature, such as sub-leading order forward-limit bounds, must be satisfied even when UV completions fall into non-localizable theories in Jaffe's language, unless momentum space Wightman functions grow too rapidly at high energy. 
Under this restriction on the growth rate, such bounds may provide IR obstructions 
to analytic, unitary, and Lorentz invariant UV completions. 
}

\begin{document}
{\baselineskip0pt
\rightline{\baselineskip16pt\rm\vbox to-20pt{
           \hbox{
KUNS-2750}
\vss}}%
}

\maketitle
\flushbottom
\section{Introduction}
Several theories have been extensively studied as a candidate for UV completed fundamental theory which describes our real world, such as string theory. 
While it must be important to seek for such theories from theoretical aspects, low energy effective field theories (LEEFT) have played an important role from the phenomenological point of view. Many LEEFT have been constructed to explain the observational data. For example, in cosmology, so-called modified gravity models have been extensively studied. 
One expectation is that it may be possible to reveal some nature of UV completion by constraining the LEEFT parameters by observations. However, unless the connection between the UV completion and LEEFT becomes clear, it would be very difficult to obtain some information of the fundamental theory from observations. From this perspective, it will be very important to investigate quantitatively how the information of UV completion is encoded in IR data.


Recently, it has been argued that the 2 to 2 scattering amplitudes of low energy effective field theories must satisfy an infinite number of inequalities, so-called positivity bounds, in order to admit a local, analytic, unitary, and Lorentz invariant UV completion with a mass gap \cite{Adams:2006sv, Bellazzini:2016xrt, deRham:2017avq}, and these bounds are applied to various models ({\it e.g.}, \cite{Bonifacio:2016wcb, deRham:2017imi, deRham:2017xox, Bellazzini:2017fep}). The existence of such bounds means that locality, unitarity, analyticity, and Lorentz invariance of UV completions are secretly encoded in LEEFT. Then, the following question will naturally arise: {\it 
all the assumptions are really necessary to derive positivity bounds obtained in the literature?} 
If one can clarify which conditions on UV completions are necessary to derive positivity bounds, one could extract information of UV completion more precisely, only from IR data. For example, if positivity bounds were successfully derived without assuming locality, such bounds could be useful to test unitarity, analyticity, or Lorentz invariance of UV completion only from IR data.  

This motivates us to investigate whether we can derive positivity bounds 
even when the locality assumption on UV completion is removed. 
It is known that the modulus of the forward scattering amplitude cannot grow faster than $s\,(\log s)^2$ in the limit $s\to\infty$ due to the unitarity bound, assuming the polynomial boundedness of the amplitude at unphysical region and existence of a mass gap. This bound is called Froissart-Martin bound \cite{Froissart:1961ux, Martin:1962rt}. Here, $s$ denotes one of the standard Mandelstam variable, corresponding to the center of mass energy. Because it is often naively argued that the locality implies the polynomial boundedness of the scattering amplitude in $s$ even in the unphysical region, the locality and unitarity are thought to be encoded in the scattering amplitude in the form of the Froissart-Martin bound. This boundedness properties lead to the 2-subtracted dispersion relation of the amplitudes, and is fully utilized in the derivation of positivity bounds \cite{Adams:2006sv, deRham:2017avq}. Then, how is the scattering amplitude bounded at high energy in non-local theories? It is expected that the non-locality will allow amplitudes to grow more rapidly than the case where locality assumption is imposed. In order to derive how the scattering amplitude is bounded at high energy in the non-local theory quantitatively, 
we consider the $2$ to $2$ massive scalar scattering in Jaffe's class of strictly localizable/quasi-local/non-localizable field theories. In such theories, the exponential growth of the 
momentum space Wightman functions can grow exponentially at high energy. Non-locality is incorporated in these theories by allowing the momentum space Wightman function to be highly singular at high energy, and it is not necessary to specify an explicit form of Hamiltonian or Lagrangian to define the non-locality. This class of theories has been well investigated: Wightman formulation of these theories has been developed \cite{Iofa:1969fj, Iofa1969}, and it is known that one can define unitary S-matrix which has standard properties as it has in standard local quantum field theories, such as cluster decomposition, LSZ construction, crossing symmetry, and CPT symmetry \cite{Steinmann1970}.

After obtaining the high energy behavior of the scattering amplitude, we investigate whether the dispersion relation with finite number of subtraction can be derived or not, and clarify under which condition positivity bounds can be obtained.

This paper is organized as follows: in sec~\ref{nonlocal}, we firstly review the definition and several important properties of Jaffe's class of strictly local/quasi-local/non-localizable theories. In sec.~\ref{sechigh}, firstly we introduce some basics on scattering amplitude. Next, we list up the assumptions we used to derive the results of this study and explain motivations to assume these conditions. Then, we derive the high-energy behavior on the forward scattering amplitude. In sec.~\ref{secposibound1}, we derive positivity bounds for $\alpha>0$ theory by making use of the results obtained in sec.~\ref{sechigh}. Sec.~\ref{concl} is devoted to conclusion and discussion of this study.
We adopt the units with $c=\hbar=1$ and the following notation:
\begin{align*}
p^2\coloneqq\eta_{\mu\nu}p^\mu p^\nu=-\left(p^0\right)^2+\delta_{ij}p^ip^j\,,\quad \eta_{\mu\nu}={\rm diag}(-1,+1,+1,+1)\,.
\end{align*}

\section{Strict localizability, quasi-locality, and non-localizability}\label{nonlocal}
In this section, we explain the definition of Jaffe's classification of quantum field theories and its physical meaning. In sec~\ref{jaffedef}, we introduce the definition of the Jaffe's classification and several essential properties of Jaffe's theories which are relevant to our work, following \cite{Keltner:2015xda}. In sec.~\ref{timeord}, we explain several important properties of the time-ordered correlation functions in Jaffe's theories.

\subsection{Definition}\label{jaffedef}
Following \cite{Keltner:2015xda}, we briefly summarize the definition of the Jaffe's classification and its relation to non-locality. In \cite{Jaffe:1966an, Jaffe:1967nb, Meiman:1964}, the criterion which classify QFT into so-called strictly localizable theories and non-localizable theories was given. This criterion was given in terms of the growth rate of the $n$-point Wightman functions in momentum space for physical time-like momenta $\{k_i\}$, which is parameterized by $\alpha$ as 
\begin{align}
\left|W\left(k_1,\cdots,k_n\right)\right|<({\rm constant})\times\left(\sum_{i=1}^n\|k_i\|\right)^{2N}\exp\left[\sigma\left(\sum_{i=1}^n\|k_i\|\right)^{2\alpha}\right]\,.\label{growthdef1}
\end{align}
Here, $\sigma$ is some positive constant and $\|X\|$ denotes the Euclidean length of a real vector $X$: $\|X\|\coloneqq\sqrt{\sum_{i=1}^4X_i^2}$. $N$ is a non-negative constant. $W\bigl(\{k_i\}\bigr)$ denotes the $n$-point Wightman function in momentum space, and trivial delta function which expresses total momentum conservation is abbreviated in the above expression. 
For this $\alpha$ parameter, fields are classified into 3 classes as
\[\begin{cases} 
0\leq\alpha<\frac{1}{2}&:\,{\rm strictly\, localizable\,field}\\
\alpha=\frac{1}{2}&:\,{\rm quasi-local\,field}\\
\alpha>\frac{1}{2}&:\,{\rm non-localizable\,field}\,.
\end{cases}
\]
Note that $\alpha=0$ case is special: in this case, the growth rate of Wightman functions is polynomially bounded for physical momenta, {\it i.e.}, Wightman functions are tempered distributions, and thus this case is especially called tempered localizable. 

In this classification, the non-local nature is naturally incorporated in the theory by considering the highly singular Wightman functions in momentum space, and it is not necessary to specify an explicit form of Hamiltonian or Lagrangian to define the non-locality. Then, why are $\alpha<\frac{1}{2}$ case, $\alpha=\frac{1}{2}$ case, and $\alpha>\frac{1}{2}$ case called strictly localizable, quasi-local, and non-localizable, respectively? 
In order to understand the physical meaning of this classification based on the value of $\alpha$, it would be instructive to consider the Lehmann-K{\"{a}}ll{\'{e}}n spectral density $\rho_\phi(\mu)$ of a scalar field $\phi$: 
\begin{align}
\rho_\phi(\mu)\sim\exp\left[\sigma\mu^\alpha\right]\,,\label{LKgrowth}
\end{align}
where $\rho_\phi(\mu)$ is defined by
\begin{align}
\sum_n\left|\bra{0}\phi(0)\ket{n}\right|^2\delta^{(4)}\left(p-p_n\right)=\frac{1}{(2\pi)^3}\Theta\left(p^0\right)\rho_\phi\left(-p^2\right)\,.\label{defLK}
\end{align}
Here, we omitted the polynomially growing factor in eq.~\eqref{LKgrowth} because it is irrelevant in the discussion below. From this definition, it turns out that $\rho_\phi(\mu)$ is a real and non-negative Lorentz scalar. $\Theta(z)$ is the Heaviside's step function. $\ket{n}$ denotes the multi-particle state belonging to the eigenstate of the momentum four-vector $P^\mu$ with eigenvalue $p_{n}^\mu$. The label $n$ includes all the labels which specify the multi-particle states with momentum eigenvalue $p_n$, and can include both continuous and discrete variables. From the spectral condition, $p_n$ satisfies $p^0_{n}\geq0$ and $-p_n^2\geq0$. Then, $\rho_\phi\,\Bigl(-p^2\Bigr)=0$ for $-p^2\leq0$.  
Let us define a $2$-point Wightman function of a scalar $\phi$ in momentum space, which is referred to as $W_{\phi}(p)$ in terms of the spectral density $\rho_\phi$ as
\begin{align}
W_\phi\left(p\right)
=(2\pi)\Theta\left(p^0\right)\rho_\phi\left(-p^2\right)\,.\label{momwight1}
\end{align}
Then, from eq.~\eqref{defLK}, Fourier transformation of $W_\phi(p)$ naively gives
\begin{align}
\int\frac{\mathrm{d}^4p}{(2\pi)^4}\,W_\phi(p)\,e^{ip(x-y)}=\sum_{n}\left|\bra{0}\phi(0)\ket{n}\right|^2e^{-ip_n(x-y)}=\bra{0}\phi(x)\phi(y)\ket{0}=W_\phi(x,y)\,,
\end{align}
for $x\neq y$. From the translation invariance, a 2-point Wightman function in position space can be written as
\begin{align}
W_\phi(x,y)&=\bra{0}\phi(x)\phi(y)\ket{0}=:W_\phi(x-y)\,,
\end{align}
and $W_\phi(z)$ may be expressed in terms of the spectral density as
\begin{align}
W_\phi(z)&=\int\frac{\mathrm{d}^4p}{(2\pi)^4}\,W_\phi(p)\,e^{ipz}\no\\
&=\int\frac{\mathrm{d}^4p}{(2\pi)^4}\Theta\left(p^0\right)\int^\infty_0\mathrm{d}\mu\,(2\pi)\delta\left(p^2+\mu\right)\rho_\phi(\mu)\,e^{ipz}=\int^\infty_0\mathrm{d}\mu\,\rho_\phi(\mu)W^{(\mu)}_{\rm free}(z)\,,\label{posiwight1}
\end{align}
where $W^{(\mu)}_{\rm free}(z)$ denotes a Wightman function for free scalar with mass square $m^2=\mu$, whose explicit form is 
\begin{align}
W^{(\mu)}_{\rm free}(z)\coloneqq\int\left.\frac{\mathrm{d}^3p}{(2\pi)^32p^0}\,e^{ipz}\right|_{p^0=\sqrt{\mu+\delta_{ij}p^ip^j}}\,.
\end{align}
From the asymptotic behavior of $W^{(\mu)}_{\rm free}(z)$ at $\mu\,\bigl|z^2\bigr|\gg1$, it turns out that the above expression for $W_\phi(z)$ with $z^2\neq0$ is convergent and well-defined for $\alpha<\frac{1}{2}$, but ill-defined for $\alpha>\frac{1}{2}$. This implies that one cannot define Wightman function in position space without smearing for $\alpha>\frac{1}{2}$. Note that one can define a position-space two-point Wightman function $W(z)$ for sufficiently large but finite $\bigl|z^2\bigr|$ when $\alpha=\frac{1}{2}$. In order to obtain a well-defined two-point Wightman function for $\alpha\geq\frac{1}{2}$, it is necessary to introduce a smeared field $\phi\,\bigl[f_{x_0}\bigr]$ centered at $x=x_0$:
\begin{align}
\phi\left[f_{x_0}\right]\coloneqq\int\mathrm{d}^4x\,\phi(x)f_{x_0}(x)\,,\label{smearfield}
\end{align}
where $f_{x_0}(x)$ denotes a smearing test function which is centered at $x=x_0$ defined by 
\begin{align}
f_{x_0}(x)\coloneqq\int\frac{\mathrm{d}^4k}{(2\pi)^4}\,\tilde f\left(k\right)e^{ik(x-x_0)}\,,
\end{align}
and $\tilde f(k)$ is an entire function in $k$. Note that $\phi\,\bigl[f_{x_0}\bigr]=\phi(x_0)$ when $\tilde f(k)=1$. Then, a position-space smeared two-point Wightman function $W_\phi\bigl(f_{x_0},g_{y_0}\bigr)$ is given by
\begin{align}
W_\phi\left(f_{x_0},g_{y_0}\right)&\coloneqq\bra{0}\phi\,\bigl[f_{x_0}\bigr]\,\phi\,\bigl[g_{y_0}\bigr]\ket{0}
=\int\frac{\mathrm{d}^4p}{(2\pi)^4}\,\Theta\left(p^0\right)\tilde f^*\left(p\right)\tilde g\left(p\right)\rho_\phi(-p^2)e^{ip(x_0-y_0)}\,.\label{smearwight}
\end{align}
Eq.~\eqref{smearwight} becomes well-defined if one chooses test functions $f$ and $g$ such that $\tilde f(p),\tilde g(p)<Ce^{-\frac{\sigma}{2}\|p\|^{2\alpha}}$ with some constant $C$. 
Although quasi-local theories or non-localizable theories have non-local nature as mentioned above, Wightman formulation of these theories has been developed \cite{Iofa:1969fj, Iofa1969}, and it is known that one can define unitary S-matrix which has standard properties as it has in standard local quantum field theories, such as cluster decomposition, LSZ construction, crossing symmetry, and CPT symmetry \cite{Steinmann1970}. For non-localizable field theories, the standard micro-causality condition which is expressed by local commutativity of fields is replaced by the asymptotic commutativity which expresses the macro-causality (detail discussions in this aspect can be found in {\it e.g.}, \cite{Soloviev:1999rv} and references therein).


\subsection{Time-ordered products}\label{timeord}
In order to evaluate the scattering amplitudes, it is necessary to know the time-ordered products of fields, or equivalently, retarded products of fields. We only discuss the time-ordered products below. For notational simplicity, let us consider the 2-point time-ordered correlation function, {\it i.e.}, the Feynman propagator firstly. The formal definition of the Feynman propagator is naively given by
\begin{align}
G_F\left(x,y\right)\coloneqq \Theta\left(x^0-y^0\right)W_\phi\left(x,y\right)+\Theta\left(y^0-x^0\right)W_\phi\left(y,x\right)\,.
\end{align}
However, the right-hand side of the above expression is ill-defined at the space-time points where the Wightman function diverges. As a result, one needs to regularize divergences appearing in the position-space Wightman functions in order to define the time-ordered products. Therefore, it is generally impossible to determine time-ordered correlation functions unambiguously from the momentum-space Wightman functions. Regularization scheme dependence will be inevitable. Then, the well-defined Feynman propagator in position space is formally given by \cite{Pfaffelhuber1972}
\begin{align}
G_F\left(x,y\right)\coloneqq g\left(-i\der_x,-i\der_y\right)\left[\Theta\left(x^0-y^0\right)W_g\left(x,y\right)+\Theta\left(y^0-x^0\right)W_g\left(y,x\right)\right]\,,
\end{align}
where $W_g(x,y)$ denotes a smeared 2-point Wightman function in position space defined by
\begin{align}
W_g\left(x,y\right)\coloneqq\int\frac{\mathrm{d}^4k_1}{\left(2\pi\right)^4}\int\frac{\mathrm{d}^4k_2}{\left(2\pi\right)^4}\,\left(2\pi\right)^4\delta^{(4)}\left(k_1+k_2\right)\frac{\tilde W_\phi\left(k_1,k_2\right)}{g\left(k_1,k_2\right)}e^{i\left(k_1x+k_2y\right)}\,. \label{regwight1}
\end{align}
Here, $\tilde W_\phi\bigl(k_1,k_2\bigr)$ is the 2-point Wightman function in momentum space which is related to $W_\phi(p)$ given in eq.~\eqref{momwight1} as $\tilde W_\phi\bigl(k,-k\bigr)=W_\phi(k)$. $g(k_1,k_2)$ denotes an entire function which is needed to make the above expression \eqref{regwight1} finite for an arbitrary set of points $(x,y)$ in Minkowski spacetime. Such $g$ is called an indicator function. Then, the Feynman propagator in momentum space is simply given by the convolution product of the 2-point Wightman function and a step function:
\begin{align}
G_F\left(k_1,k_2\right)&\coloneqq\int\mathrm{d}^4x_1\int\mathrm{d}^4x_2\,G_F\left(x_1,x_2\right)e^{-ik_1x_1-ik_2x_2}\no\\
&=\left(2\pi\right)^4\delta^{(4)}\left(k_1+k_2\right)g\left(k_1,k_2\right)\int^\infty_{-\infty}\frac{\mathrm{d}p^0}{2\pi}\,\left.\frac{-i}{g\left(p,-p\right)}\left[\frac{\tilde W_\phi\left(p,-p\right)}{p^0-k_1^0-i\epsilon}-\frac{\tilde W_\phi\left(-p,p\right)}{p^0-k_1^0+i\epsilon}\right]\right|_{\vec p=\vec k_1}\,.\label{Feynmanprop2}
\end{align}
Here, we chose a permutation invariant indicator function which satisfies $g(k,-k)=g(-k,k)$.\footnote{As we will mention soon, on-shell quantities such as scattering amplitudes is independent of a particular choice of an indicator.} If we choose a Lorentz invariant indicator function $g$ and write $g\left(-k^2\right)\coloneqq g\left(k,-k\right)$, and using eq.~\eqref{momwight1} with $W_\phi(p)$, the above expression of the Feynman propagator in momentum space is reduced to the well-known Lehmann-K{\"{a}}ll{\'{e}}n spectral representation by changing the integration variable in eq.~\eqref{Feynmanprop2} from $p^0$ to $\mu\coloneqq\bigl(p^0\bigr)^2-|\vec k|^2$:
\begin{align}
G_F\left(-k^2\right)=g\left(-k^2\right)\int^\infty_0\mathrm{d}\mu\,\frac{\rho_\phi(\mu)}{g(\mu)}\frac{-i}{k^2+\mu-i\epsilon}\,,\label{Feynmanprop1}
\end{align}
where $G_F\Bigl(-k^2\Bigr)$ is defined by $G_F\left(k_1,k_2\right)=\left(2\pi\right)^4\delta^{(4)}\left(k_1+k_2\right)G_F\Bigl(-k_1^2\Bigr)$. 
When $\rho_\phi(\mu)\sim \mu^N\exp\bigl[\sigma\mu^\alpha\bigr]$, we choose $g(\mu)$ which behaves as $g(\mu)\sim\mu^{N+1}\exp\bigl[\sigma\mu^\alpha\bigr]$. Then, eq.~\eqref{Feynmanprop1} is well-defined. 
Feynman propagator depends on the choice of an indicator function. Such ambiguities are expressed as contact terms, which may become manifest if we rewrite eq.~\eqref{Feynmanprop1} as 
\begin{align}
G_F\left(-k^2\right)&=\int^\infty_0\mathrm{d}\mu\,\frac{(-i)\,\rho_\phi(\mu)}{k^2+\mu-i\epsilon}\left(1-\frac{g(\mu)-g\left(-k^2\right)}{g(\mu)}\right)\no\\
&=\int^\infty_0\mathrm{d}\mu\,\frac{(-i)\,\rho_\phi(\mu)}{k^2+\mu-i\epsilon}\left[1-\frac{1}{g(\mu)}\sum_{n=1}^{\infty}\frac{1}{n!}\frac{\der^n\,g\left(-k^2\right)}{\der\left(-k^2\right)^n}\left(\mu+k^2\right)^n\right]\,.
\end{align}
The second term in the final line expresses the divergent contact terms. These contact terms cancel on-shell singularity included in the Feynman propagator, and hence do not contribute to on-shell quantities. When $\alpha=0$, namely, the theory is tempered-localizable, an indicator $g(\mu)$ is an real polynomial, and hence the number of contact terms is finite. When $\alpha>0$, an indicator $g(\mu)$ is given by an entire function, and hence the number of contact terms is infinite. We do not expect the appearance of an infinite number of contact terms within a validity of a perturbation theory. Therefore, quantum field theories whose high-energy behavior of them can be captured perturbatively will fall into $\alpha=0$ theory. However, the appearance of an infinite number of contact terms does not necessarily mean that the number of parameters included in the theory is infinite. Thus, it seems that even the perturbatively renormalizable theory can be $\alpha>0$ theory generically, although in $\alpha>0$ case it will be difficult to determine the value of $\alpha$ for some given QFT rigorously, because in order to do so one needs to evaluate the high-energy behavior of the spectral density $\rho(\mu)$ non-perturbatively.\footnote{It is also expected from perturbative analysis that an infinite number of divergences will appear in perturbatively nonrenormalizable theories. The connection between perturbatively nonrenormalizable theories and non-localizable theories was discussed in \cite{Schroer:1964}.} There are several theories which are conjectured to be $\alpha>0$ theories. For example, it has been conjectured that Little String Theories may be $\alpha=\frac{1}{2}$ quasi-local theories in \cite{Kapustin2011}, and  Galileon theories may be $\alpha>\frac{1}{2}$ non-localizable theories in \cite{Keltner:2015xda}. 
 
In fact, eq.~\eqref{Feynmanprop1} is also useful for investigating the behavior of the Feynman propagator for large modulus of $-k^2\in\mathbb C$. This is because $g\Bigl(-k^2\Bigr)$ is an entire function and hence $G_F\Bigl(-k^2\Bigr)$ is analytic in $-k^2$ in the complex $-k^2$-plane modulo an isolated pole and branch cut. 
Therefore, one can easily analytically continue $G_F\Bigl(-k^2\Bigr)$ for $-k^2\in\mathbb C$ and it turns out that the growth rate of the Feynman propagator $G_F\Bigl(-k^2\Bigr)$ for large modulus of $-k^2\in\mathbb C$ is bounded as\footnote{Note that the left-hand side of \eqref{Feynmanbound1} will be non-zero for time-like $k$, while it can be zero for space-like $k$, generally. This is because an indicator $g\,\Bigl(-k^2\Bigr)$ must grow for large $-k^2$, while it may not grow for different directions in a complex $-k^2$-plane. In some special case $g\,\Bigl(-k^2\Bigr)$ can decay rapidly for Euclidean direction. The rapid decay of $g\,\Bigl(-k^2\Bigr)$ in Euclidean direction is assumed and used as a key property regulating perturbative UV loops in Efimov's non-local QFT program (see {\it e.g.}, \cite{Efimov1967}), although this rapid decay might not necessarily be ensured generically in quasi-local/ non-localizable theories in Jaffe's language, as is also pointed out in \cite{Keltner:2015xda}.}
\begin{align}
\lim_{\left|-k^2\right|\to\infty}\left|\frac{G_F\left(-k^2\right)}{\left(-k^2\right)^{N}e^{\sigma\left|-k^2\right|^{\alpha}}}\right|<\infty\,,\label{Feynmanbound1}
\end{align}
when $-k^2$ is not on singularities. This boundedness property obeys from the growth rate of an indicator function via eq.~\eqref{Feynmanprop1}. This means that the growth rate of the Feynman propagator for large  modulus of $-k^2\in\mathbb C$ is determined by the growth rate of the 2-point Wightman function $W_\phi(k)$ at large $-k^2$ with {\it physical time-like} $k$. It is not necessary to specify the properties of the 2-point Wightman function $W_\phi(k)$ at large modulus of $-k^2$ with unphysical $k$ to derive eq.~\eqref{Feynmanprop1}. 

Next, let us generalize the above discussion to the $4$-point time-ordered correlation function in momentum space, because we will consider the $2$ to $2$ scattering amplitude which is related to $4$-point time-ordered correlation functions through the reduction formula. As in the case of the $2$-point function, the $4$-point time ordered correlation function is also formally defined by 
\begin{align}
&G_F\left(x_1,\cdots,x_4\right)\no\\
&\coloneqq g\left(-i\der_{x_1},\cdots,-i\der_{x_4}\right)\left[\sum_I\Theta\left(x_{i_1}^0-x_{i_2}^0\right)\Theta\left(x_{i_2}^0-x_{i_3}^0\right)\Theta\left(x_{i_3}^0-x_{i_4}^0\right)W_g\left(x_{i_1},\cdots,x_{i_4}\right)\right]\,,\label{regFeyn2}
\end{align}
where $I$ denotes the permutations $\Bigl(\begin{smallmatrix}
1,&\cdots,&4 \\ 
i_1,&\cdots,&i_4 \end{smallmatrix}\Bigr)$, and $\sum_I$ denotes the summation over all permutations. Here, $W_g(x_1,\cdots, x_4)$ is a smeared 4-point Wightman function defined by
\begin{align}
W_g\left(x_1,\cdots,x_4\right)\coloneqq\left[\prod_{i=1}^4\int\frac{\mathrm{d}^4k_i}{(2\pi)^4}\right](2\pi)^4\delta^{(4)}\left(\sum_{j=1}^4k_j\right)\frac{\tilde W_\phi\left(k_1,\cdots, k_4\right)}{g\left(k_1,\cdots, k_4\right)}\,e^{i\sum_{n=1}^4k_nx_n}\,,
\end{align}
and we chose the permutation-invariant indicator function in eq.~\eqref{regFeyn2} for simplicity:
\begin{align}
g\left(k_1,\cdots,k_4\right)=g\left(k_{i_1},\cdots,k_{i_4}\right)\quad{\rm for}\,\,\,\forall I\,.
\end{align}
Then, momentum space 4-point time-ordered correlation function is simply given by the convolution product of $4$-point Wightman function and step functions:
\begin{align}
G_F\left(k_1,\cdots, k_4\right)&\coloneqq\left[\prod_{i=1}^4\int\mathrm{d}^4x_i\right]\,G_F\left(x_1,\cdots,x_4\right)e^{-i\sum_{j=1}^4k_jx_j}\no\\
&=(2\pi)^4\delta^{(4)}\left(\sum_{i=1}^4k_i\right)g\left(k_1,\cdots ,k_4\right)\sum_{I}\left[\prod_{j=1}^4\int\frac{\mathrm{d}p_j^0}{2\pi}\right](2\pi)\delta\left(\sum_{l=1}^4p_l^0\right)\no\\
&\quad\times\left.\frac{\tilde W_\phi\left(p_{i_1},\cdots, p_{i_4}\right)}{g\left(p_{i_1},\cdots, p_{i_4}\right)}\right|_{\left\{\vec p_{i_n}=\vec k_{i_n}\right\}_{n=1,\cdots,4}}\prod_{r=1}^{3}\left(\frac{-i}{\sum_{n=1}^r\left(p_{i_n}^0-k_{i_n}^0\right)-i\epsilon}\right)\,.\label{Feynmanprop3}
\end{align}
This is just the generalization of eq.~\eqref{Feynmanprop2} to the $4$-point case.
Then, using the fact that $4$-point Wightman function can be written in terms of a Lorentz scalar function $\mathcal W_I$ as 
\begin{align}
&\left.\tilde W_\phi\left(p_{i_1},\cdots, p_{i_4}\right)\right|_{p_1+\cdots+p_4=0}=\left[\prod_{n=1}^3\Theta\left(p_{i_n}^0\right)\right]\left.\mathcal{W}_I\left(-\left(p_1+p_2\right)^2,\,-\left(p_1-p_3\right)^2,\,\left\{p_i^2\right\}\right)\right|_{p_1+\cdots+p_4=0}\,,
\end{align}
and choosing a Lorentz invariant indicator function $g=g\Bigl(-\bigl(k_1+k_2\bigr)^2,\,-\bigl(k_1-k_3\bigr)^2,\,\bigl\{k_i^2\bigr\}\Bigr)$, eq.~\eqref{Feynmanprop3} reduces to
\begin{align}
&\frac{G_F\left(-\left(k_1+k_2\right)^2,\,-\left(k_1-k_3\right)^2,\,\left\{k_i^2\right\}\right)}{g\left(-\left(k_1+k_2\right)^2,\,-\left(k_1-k_3\right)^2,\,\left\{k_i^2\right\}\right)}\no\\
&=\sum_I\int^\infty_{-|\vec k_1+\vec k_2|^2}\frac{\mathrm{d}s'}{2\pi}\int\frac{\mathrm{d}\left(p_1^0-p_2^0\right)}{4\pi}\int\frac{\mathrm{d}p_3^0}{2\pi}\left.\frac{\mathcal{W}_I\left(s',\,-\left(p_1-p_3\right)^2,\,\left\{p_i^2\right\}\right)}{g\left(s',\,-\left(p_1-p_3\right)^2,\,\left\{p_i^2\right\}\right)}\right|_{\left\{\vec p_{i_n}=\vec k_{i_n}\right\}_{n=1,\cdots,4},\,p_4^0=-\sum_{j=1}^3p_j^0}\no\\
&\quad\times\sum_{\lambda=\pm1}\left.\frac{\Theta\left(\lambda\left(p_1^0+p_2^0\right)\right)}{2\lambda\left(p_1^0+p_2^0\right)}\prod_{r=1}^{3}\left(\frac{-i}{\sum_{n=1}^r\left(p_{i_n}^0-k_{i_n}^0\right)-i\epsilon}\,\Theta\left(p_{i_r}^0\right)\right)\right|_{p_1^0+p_2^0=\lambda\left(s'+|\vec k_1+\vec k_2|^2\right)^{\frac{1}{2}}}\,,\label{Feynmanprop4}
\end{align}
where
\begin{align}
&G_F\left(k_1,\cdots, k_4\right)\coloneqq(2\pi)^4\delta^{(4)}\left(k_1+\cdots+k_4\right)G_F\left(-\left(k_1+k_2\right)^2,\,-\left(k_1-k_3\right)^2,\,\left\{k_i^2\right\}\right)\,.
\end{align}
Here, $\bigl\{k_i^2\bigr\}$ stands for $\bigl\{k_i^2\bigr\}_{i=1,\cdots,4}$. In order to obtain eq.~\eqref{Feynmanprop4}, we changed the integration variables by introducing $s'\coloneqq\bigl(p_1^0+p_2^0\bigr)^2-|\vec k_1+\vec k_2|^2$. Again, $G_F(k_1,\cdots,k_4)$ can be determined only up to contact terms, although they do not contribute to scattering amplitudes. As in the case of the $2$-point Feynman propagator, it is expected from eq.~\eqref{Feynmanprop4} that the $4$-point time-ordered correlation functions in momentum space will be also bounded for large modulus of $-\bigl(k_1+k_2\bigr)^2\in\mathbb C$ with fixed $-\bigl(k_1-k_3\bigr)^2$ and $\bigl\{k_i^2\bigr\}_{i=1,\cdots,4}$ as 
\begin{align}
\lim_{\left|-\left(k_1+k_2\right)^2\right|\to\infty}\left|\frac{G_F\left(-\left(k_1+k_2\right)^2,\,-\left(k_1-k_3\right)^2,\,\left\{k_i^2\right\}\right)}{\left(-\left(k_1+k_2\right)^2\right)^{N}e^{\sigma'\left(-\left(k_1+k_2\right)^2\right)^{\alpha}}}\right|<\infty\,,\label{Feynmanbound2}
\end{align}
when 
the arguments of $G_F$ are not on singularities. 
Here, $\sigma'$ is some positive constant whose precise value of $\sigma'$ may depend on the value of $-(k_1-k_3)^2$, $\bigl\{k_i^2\bigr\}$, or $\sigma$ appearing in eq.~\eqref{growthdef1},  but the precise value of $\sigma'$ is irrelevant in our study.   
It should be noted in advance that this boundedness property \eqref{Feynmanbound2} of time-ordered correlation functions is an important observation which would support the plausibility of the assumption C introduced in the next sec.~\ref{sechigh}.

\section{Bounds on the high energy behavior of scattering amplitudes for physical $s$}\label{sechigh}
In this section, we will obtain the high energy behavior of the forward-limit scattering amplitude under several assumptions. 
In sec.~\ref{assumption1}, we firstly introduce some basics of $2$ to $2$ scattering amplitudes. Next, we list up the assumptions which are needed to obtain the high energy behavior of the scattering amplitude and the dispersion relation. We also explain the physical meaning or motivations of these assumptions. In sec.~\ref{subsechigh}, we derive the high energy behavior of the scattering amplitude for $\alpha\geq0$ case. We also explain the intuitive derivation of the high energy behavior.
\subsection{Basics and assumptions}\label{assumption1}
Let us consider the $2$ to $2$ scattering of the real scalar $\phi$ with positive mass square $m^2>0$. We refer to the incoming momenta and outgoing momenta as $p_1,p_2$ and $p_3,p_4$, respectively. From the total momentum conservation, $p_4=p_1+p_2-p_3$. Then, from the Lorentz invariance, one can express the corresponding scattering amplitudes in terms of the Mandelstam variables $s$, $t$ and $u$ which are defined by
\begin{align}
s&\coloneqq-\left(p_1+p_2\right)^2=-(p_3+p_4)^2\,,\\
t&\coloneqq-\left(p_1-p_3\right)^2=-(p_2-p_4)^2\,,\\
u&\coloneqq-\left(p_1-p_4\right)^2=-(p_2-p_3)^2\,.
\end{align}
From this definition, it turns out that $s+t+u=4m^2$ holds, and hence we refer to the scattering amplitude of this process by $F(s,t)$. In the center of mass frame (CM frame), on-shell momenta $p_1$, $p_2$, and $p_3$ can be parameterized as 
\begin{align}
&p^\mu_1=\left(\frac{E_{\rm cm}}{2},0,0,q_s\right)\,,\quad p^\mu_2=\left(\frac{E_{\rm cm}}{2},0,0,-q_s\right)\,,\quad p_3^\mu=\left(\frac{E_{\rm cm}}{2},q_s\sin\theta,0,q_s\cos\theta\right)\,.
\end{align}
Here, $q_s=\frac{1}{2}\sqrt{s-4m^2}$ denotes the amplitude of the three-momentum in the CM frame. Then, Mandelstam variables can be written in terms of the 4-momenta in CM frame as
\begin{align}
s=E^2_{\rm cm}\,,\quad t=2q_s^2\left(\cos\theta-1\right)\,,\quad u=-2q_s^2\left(\cos\theta+1\right)\,,
\end{align}
and so $s$ and $t$ are called center of mass energy and momentum transfer, respectively.

From now on, in order to make the discussion clearer, we list up the assumptions which are needed to obtain the results presented in this paper:
\begin{description}
\item{A:} Lorentz invariance, unitarity, and $s\leftrightarrow t\leftrightarrow u$ crossing symmetry

\item{B1: Analyticity properties in the large Lehmann ellipse}\\
For fixed $s\in \mathbb{R}+i\epsilon$ for sufficiently large $\mathrm{Re}\, s$ with an infinitesimal positive constant $\epsilon$, $\tilde F(s,z)\coloneqq F(s,t)\bigr|_{t=2q_s^2(z-1)}$ is assumed to be holomorphic in $z$ on or inside the large Lehmann ellipse $\mathcal C_0$ with foci at $z=\pm1$, and the semi-major axis is $1+\bigl(t_0/2q_s^2\bigr)$. The ellipse $\mathcal C_0$ can be expressed as $\mathcal C_0=\bigl\{z'=z(s)\!:\, z(s)\!\coloneqq\cosh\bigl(\beta_0(s)+i\varphi'\bigr)\,, -\pi<\varphi'\leq\pi\bigr\}$ with $\cosh\bigl(\beta_0(s)\bigr)=1+\bigl(t_0/2q_s^2\bigr)$. Here, $t_0>0$ is an $s$-independent constant. We expect that we can take $t_0=m^2(1-\delta)$ with $0<\delta\ll1$, although the specific value of $t_0$ does not change any discussions below. 
\item{B2: Analyticity properties in the complex $s$-plane}\\
The forward scattering amplitude $F(s,0)$ is assumed to be holomorphic in $s$ in the first Riemann sheet of the complex $s$-plane modulo poles and branch cuts which are predicted by unitarity. $F(s,t)$ is assumed to be analytic in $t$ at $t=0$ for any $s$ in the first Riemann sheet of the complex $s$-plane modulo poles and branch cuts.
\item{C: Exponential boundedness of the amplitude for fixed $t$ at large $s$}\\
On the ellipse $\mathcal C_0$, $\tilde F\bigl(s,z(s)\bigr)$ is assumed to be exponentially bounded:\footnote{This boundedness property in the complex $z$-plane has been already obtained in \cite{Fainberg1971} for $\alpha>\frac{1}{2}$ case. 
}
\begin{align}
\left|\tilde F\left(s,z(s)\right)\right|<\tilde Cs^N\exp\left[Cs^\alpha\right]\,,\quad N<\infty\,,\quad 0<C\,,\tilde C\,,\quad {\rm as }\quad s\to\infty+i\epsilon\,.\label{expbound0}
\end{align}
Furthermore, we also assume that $\left|\der^n_t F(s,t)\right|_{t=0}$ is also exponentially bounded in the complex $s$-plane: 
\begin{align}
\left|\der^n_t F(s,t)\right|_{t=0}<\tilde C_n\,|s|^{N_n}\exp\left[C_n|s|^\alpha\right]\,,\quad N_n<\infty\,,\quad 0<C_n\,,\tilde C_n\,,\quad {\rm as}\quad |s|\to\infty\,.\label{expbound2}
\end{align}
Here, $C$, $\tilde C$, $C_n$, and $\tilde C_n$ denote some positive constants.
\end{description}
\subsubsection*{Motivations for assuming conditions B1-C}
There are several motivations why it would be meaningful to work under the assumptions B1-C. 
\begin{itemize}
\item Motivation for assuming B1: \\
Assumption B1 plays an essential role to constrain the high energy behavior of the scattering amplitude. As we will discuss at the end of sec.~\ref{subsechigh}, the holomorphy in the ellipse $\mathcal C_0$ is closely related to the short-range nature of the force which induces the scattering under consideration.\footnote{This point is also mentioned in \cite{Martin:1969ina}.} We expect this holomorphy in the ellipse $\mathcal C_0$ as long as the scatterings are caused by a short-range force. 
\item Motivation for assuming B2: \\
It is known that causality and analyticity of the forward scattering amplitudes in complex $s$ plane 
are closely connected to each other, and their connection has been extensively investigated from 1950's ({\it e.g.}, \cite{PhysRev.95.1612}). Roughly speaking, the retarded propagator associated with the scattering process is given by the Fourier component of the forward scattering amplitude, and hence analyticity in the first Riemann sheet of the complex $s$-plane implies causality \cite{Giddings:2007qq, Giddings:2009gj, Keltner:2015xda}.\footnote{It is argued in \cite{Keltner:2015xda} that analyticity with allowing exponential growth \eqref{expbound2} implies macro-causality.} 
We simply assume that these properties hold for generic $\alpha$ because of its close connection to the causality. 

\item Motivation for assuming C:\\
$F(s,t)$ is written in terms of 4-point time-ordered correlation functions $G_F\bigl(p_1,p_2,-p_3, -p_4=-(p_1+p_2-p_3)\bigr)$ via the reduction formula. 
Thus, we expect that the boundedness property \eqref{Feynmanbound2} of $G_F$ in the complex $-\bigl(p_1+p_2\bigr)^2$-plane with fixed $-\bigl(p_1-p_3\bigr)^2$ and $\bigl\{p_i^2\bigr\}$ will be intimately connected to the boundedness properties of $F(s,t)$ for large $|s|$ in the unphysical region with fixed $t$.\footnote{In analytically continuing $F(s,t)$ to the unphysical region, one must maintain the on-shell condition $k_i^2=-m^2$. 
} 
Then, from \eqref{Feynmanbound2}, it is expected that $F(s,t)$ would be also bounded exponentially for large $|s|$ as \eqref{expbound0} or \eqref{expbound2}.  
 \\
\end{itemize}
Although the rigorous derivation of analyticity properties B1-2 are discussed in the context of axiomatic QFT (see {\it e.g.}, \cite{Hepp:1964, Martin:1965jj, Epstein:1969bg, Martin:1969ina}), we simply assume these properties as a starting point of this study. From now, we move on to derive several consequences followed from these assumptions. 
\subsection{High energy behavior of the scattering amplitude}\label{subsechigh}

Firstly, we develop the method to obtain a high energy behavior of the scattering amplitude at $t=0$ and its $n$-th derivative with respect to $t$ at $t=0$.  That is, we obtain the upper bound on $F^{(n)}(s)$ at large $s$, where $F^{(n)}(s)$ is defined by
\begin{align}
F^{(n)}(s)\coloneqq\left.\frac{1}{n!}\frac{\der^n}{\der t^n}F(s,t)\right|_{t=0}\,, \label{taylor1}
\end{align} 
under the assumptions $A-C$. Note that the method presented in this subsection \ref{subsechigh} is developed based on \cite{Azimov:2011nk}, although the bounds on the $F^{(n)}(s)$ with $n\geq1$ was not discussed there.\\
Partial wave expansion of $F(s,t)$ is given by
\begin{align}
F(s,t)=\frac{\sqrt{s}}{q_s}\sum_{l=0}^\infty(2l+1)f_l(s)P_l\left(1+\frac{t}{2q_s^2}\right)\,\label{partialwave1}
\end{align}
where $P_l(z)$ denotes the Legendre polynomial.
From the S-matrix unitarity $SS^\dagger=1$, one obtains the unitarity constraints on the partial wave amplitude $f_l(s)$ for $s\geq4m^2$:
\begin{align}
|f_l(s)|\leq1\,.\label{unitarypartial}
\end{align}
From the holomorphy of $\tilde F(s,z)$ on or inside the ellipse $\mathcal C_0$ 
$f_l(s)$ can be expressed as
\begin{align}
f_l(s)=\frac{q_s}{2\pi i\sqrt s}\oint_{\mathcal{C}_0} \tilde F(s,z')Q_l(z')\mathrm{d}z'\,.
\end{align}
This is known as the Neumann's expansion.
Here, $Q_l(z)$ denotes the Legendre functions of the 2nd kind. By making use of the asymptotic behavior of $Q_l(z)$ at large $l$, one obtains
\begin{align}
|f_l(s)|<G_l(\beta_0,s)\coloneqq\frac{q_s}{\sqrt{s}}B_0(s)\sqrt{e^{-\beta_0}\cosh\beta_0}\cdot\frac{e^{-\beta_0l}}{\sqrt l}\,,\label{shortbound}
\end{align}
where
\begin{align}
B_0(s)\coloneqq\frac{1}{\sqrt{8\pi}}\int^\pi_{-\pi}\mathrm{d}\varphi'\left|\tilde F\left(s,\cosh\left(\beta_0(s)+i\varphi'\right)\right)\right|\,.\label{B0def}
\end{align}
Defining $L(s)$ by $G_L<1\leq G_{L-1}$, one obtains 
\begin{align}
G_{l'+L}(\beta_0,s)=\frac{e^{-\beta_0l'}}{\sqrt{1+\frac{l'}{L}}}G_L(\beta_0,s)<\frac{e^{-\beta_0l'}}{\sqrt{1+\frac{l'}{L}}}\,.\label{Ldef}
\end{align}
Thanks to the bound \eqref{shortbound} and from the asymptotic behavior of the Legendre polynomial $P_l(z)$ in the limit $l\to\infty$, Weierstrass's theorem ensures that partial wave expansion \eqref{partialwave1} uniformly converges inside the ellipse $\mathcal C_0$ when $s\in\mathbb{R}+i\epsilon$. Because $P_l(z)$ is holomorphic in the whole complex $z$-plane, Weierstrass's double series theorem ensures that the $n$-th derivative of $F(s,z)$ with respect to $z$ is given by
\begin{align}
\frac{\der^n}{\der z^n}F(s,z)=\frac{\sqrt s}{q_s}\sum_{l=0}^\infty(2l+1)f_l(s)P^{(n)}_l(z)\,,\quad P^{(n)}_l(z)\coloneqq\frac{\der^n}{\der z^n}P_l(z)\,,
\end{align}
inside the ellipse $\mathcal C_0$.
Then, when $s\in\mathbb{R}+i\epsilon$, the $n$-th derivative of $F(s,t)$ with respect to $t$ at $t=0$ is given by
\begin{align}
F^{(n)}(s)=\frac{\sqrt s}{q_s}\frac{1}{n!2^nq_s^{2n}}\sum_{l=0}^\infty(2l+1)f_l(s)P^{(n)}_l(1)\,.\label{diffpartialwave1}
\end{align}


We have obtained two constraints on $|f_l(s)|$ so far: \eqref{unitarypartial} and \eqref{shortbound}. From the definition of $L(s)$, it turns out that \eqref{unitarypartial} and \eqref{shortbound} give the stronger bound for $l\leq L(s)-1$ and for $l\geq L(s)$, respectively.
Therefore, by using \eqref{Ldef},
\begin{align}
\left|F^{(n)}(s)\right|&<\frac{\sqrt s}{q_s}\frac{1}{n!2^nq_s^{2n}}\left[\sum_{l=0}^{L-1}(2l+1)P^{(n)}_l(1)+\sum_{l=L}^\infty(2l+1)G_{l}(\beta_0,s)P^{(n)}_l(1)\right]\no\\
&<\frac{\sqrt s}{q_s}\frac{1}{n!2^nq_s^{2n}}\left[\sum_{l=0}^{L-1}(2l+1)P^{(n)}_l(1)+\sum_{l'=0}^\infty e^{-\beta_0l'}P^{(n)}_{l'+L}(1)\left(2L\left(1+\frac{l'}{L}\right)^{\frac{1}{2}}+\left(1+\frac{l'}{L}\right)^{-\frac{1}{2}}\right)\right]\,.\label{diffbound}
\end{align}
Because
\begin{align}
P_l(z)={}_2F_1\left(l+1,-l;1;\frac{1-z}{2}\right)\,,
\end{align}
and Gauss's hyper geometric function ${}_2F_1\left(\alpha,\beta;\gamma;x\right)$ satisfies
\begin{align}
\frac{\der}{\der x}\,{}_2F_1\left(\alpha,\beta;\gamma;x\right)=\frac{\alpha\beta}{\gamma}{}_2F_1\left(\alpha+1,\beta+1;\gamma+1;x\right)\,,
\end{align}
$P^{(n)}_l(1)$ can be evaluated as
\begin{align}
P^{(n)}_l(1)=\begin{cases}
\frac{1}{2^n}\frac{(l+n)!}{(l-n)!} & n\leq l\,,\\
0 & n\geq l+1\,. \label{difflegendre}
\end{cases}
\end{align}
Substituting eq.~\eqref{difflegendre} into eq.~\eqref{diffbound},
\begin{align}
\left|F^{(n)}(s)\right|
&<\frac{\sqrt s}{q_s}\frac{1}{n!4^nq_s^{2n}}\Biggl[\sum_{l=n}^{L-1}(2l+1)l^{2n}\left(1+\mathcal{O}\left(l^{-1}\right)\right)\no\\
&\hspace{1 cm}+\sum_{l'=0}^\infty e^{-\beta_0l'}(l'+L)^{2n}\left(1+\mathcal{O}\left(\left(l'+L\right)^{-1}\right)\right)\left(2L\left(1+\frac{l'}{L}\right)^{\frac{1}{2}}+\left(1+\frac{l'}{L}\right)^{-\frac{1}{2}}\right)\Biggr]\,,\no\\
&\sim\frac{\sqrt s}{q_s}\frac{1}{q_s^{2n}}\left[L^{2n+2}+\frac{2L^{\frac{1}{2}}}{\beta_0}\int^\infty_0\mathrm{d}y\,e^{-y}\frac{(y+Y)^{2n+\frac{1}{2}}}{\beta_0^{2n+\frac{1}{2}}} \right]\,,
\label{diffbound2}
\end{align}
with neglecting the irrelevant proportionality constant. Here, $y\coloneqq \beta_0l'$ and $Y\coloneqq \beta_0L$, and we assumed that $L$ and $Y$ grows as $s$ increases and neglected the terms which are lower order in $L$ or $Y$. When $L$ grows as $s$ increases, $L$ and $B_0(s)$ are related as
\begin{align}
B_0(s)\simeq2e^{\beta_0L}\sqrt{L}\,,\label{L-Brel}
\end{align}
which can be obtained from the definition of $L$.
\subsubsection*{\textbullet\,\,\,\,$\alpha=0$ case}
By using eqs.~\eqref{B0def} and \eqref{L-Brel}, polynomial boundedness \eqref{expbound0} with $\alpha=0$ is converted to the bound on $Y=\beta_0L$ as
\begin{align}
Y<\left(N-\frac{1}{4}\right)\ln(s/s_0)-\frac{1}{2}\ln Y\approx\left(N-\frac{1}{4}\right)\ln(s/s_0)-\frac{1}{2}\ln\ln(s/s_0)\,,\label{Ybound1}
\end{align}
where $s_0$ is some constant. Applying eq.~\eqref{Ybound1} to eq.~\eqref{diffbound2}, and by making use of the approximation $2\beta_0\approx(t_0/s)^{1/2}$ which is valid
at large $s$, one obtains
\begin{align}
\left|F^{(n)}(s)\right|<\left({\rm constant}\right)\times s\left(\ln(s/s_0)\right)^{2n+2}\,.\label{frois1}
\end{align}
$n=0$ case of eq.~\eqref{frois1} gives the bound on the total cross section $\sigma_{\rm tot}(s)$ as 
\begin{align}
\sigma_{\rm tot}(s)<\left({\rm constant}\right)\times \left(\ln(s/s_0)\right)^2\,,
\end{align}
which corresponds to a celebrated Froissart-Martin bound\footnote{More precisely, the bound on $\sigma_{\rm tot}(s)$ which can be obtained from this analysis is slightly stronger than a simple log-square form, because of the second term in eq.~\eqref{Ybound1}. This fact is pointed out in \cite{Azimov:2011nk}.} \cite{Froissart:1961ux, Martin:1962rt}. Eq.~\eqref{frois1} naively suggests that $s^{-2}\cdot\bigl|F(s,t)\bigr|\to0$ as $s\to\infty+i\epsilon$ for small but positive $t$. Indeed, the bound on $F(s,t)$ obtained in \cite{Azimov:2011nk} for $t>0$ also implies the existence of such small but positive $t$ if we assume the polynomial boundedness \eqref{expbound0} with $\alpha=0$. This is in fact crucially important to obtain a dispersion relation of $F(s,t)$ for $t>0$ with finite number of subtractions \cite{Jin:1964zza}.

\subsubsection*{\textbullet\,\,\,\,$\alpha>0$ case}\label{highbe2}
Next, let us derive the bound on $\left|F^{(n)}(s)\right|$ at $s\to\infty$ assuming the exponential boundedness \eqref{expbound0} with $\alpha>0$. Then, from eq.~\eqref{B0def}, $B_0(s)$ is also bounded as
\begin{align}
B_0(s)<\tilde C\sqrt{\frac{\pi}{2}}s^N\exp\left[Cs^\alpha\right]\,.
\end{align}
From this bound and eq.~\eqref{L-Brel}, $Y=\beta_0L$ is bounded as
\begin{align}
Y<Cs^\alpha+\left(N-\frac{1}{4}\right)\ln(s/s_0)-\frac{1}{2}\ln Y\approx Cs^\alpha+\left(N-\frac{\alpha}{2}-\frac{1}{4}\right)\ln(s/s_0)-\frac{1}{2}\ln\ln(s/s_0)\,.\label{Ybound2}
\end{align}
Applying the bound \eqref{Ybound2} to eq.~\eqref{diffbound2}, for $n\geq0$, one obtains
\begin{align}
\left|F^{(n)}(s)\right|<\frac{C^{2n+2}s^{1+(2n+2)\alpha}}{{t_0}^{n+1}}
\,.\label{realsbound1}
\end{align}
Note that the right-hand side diverges in the limit $t_0\to0$, which is consistent with the fact that $|F(s,0)|$ diverges when the massless particle exchange exists. This result shows that $|F(s,0)|$ is {\it polynomially bounded} in $s$ when $s\to\infty+i\epsilon$, even if $\bigl|F(s,t)\bigr|$ can grow exponentially in the unphysical region. The powers of $s$ of the right-hand side of \eqref{realsbound1} grows as $n$ grows, however. This suggests that $\bigl|F(s,t)\bigl|$ grows faster than any polynomial in $s$ in the Regge limit when $t>0$. This is consistent with the assumption C which we imposed to derive this result, and indeed, the bound on $F(s,t)$ obtained in \cite{Azimov:2011nk} for $t>0$ also implies the exponential growth of $F(s,t>0)$ in the Regge limit if we assume the exponential growth \eqref{expbound0} with $\alpha>0$. This result implies that one cannot obtain dispersion relation with finite number of subtractions for $F(s,t)$ when $t>0$. \eqref{realsbound1} also suggests that the Froissart-Martin bound can be violated even in local QFT with $0<\alpha<\frac{1}{2}$. We do not deny the possibility that $\tilde F(s,z)$ were in fact polynomially bounded in $s$ on or inside the ellipse $\mathcal C_0$ even when $0<\alpha<\frac{1}{2}$, but we do not expect such boundedness property from the consideration in sec.~\ref{timeord}. It should be noted that the forward scattering amplitude $F(s,0)$ is still bounded by $s^2$ for $0\leq\alpha<\frac{1}{2}$ in spite of the possible violation of the Froissart-Martin bound for $\alpha>0$:
\begin{align}
\lim_{s\to\infty+i\epsilon}\left|\frac{F(s,0)}{s^2}\right|=0\quad {\rm for}\quad 0\leq\alpha<\frac{1}{2} \,.
\end{align}

Intuitively, the origin of the existence of the bounds \eqref{shortbound} on the partial wave amplitudes $f_l(s)$ can be understood as the short-range nature of the force mediating the scattering under consideration. In the large impact parameter regime $b\sim(l/q_s)\gg1$ 
and for fixed $0\leq-t\ll s$, {\it i.e.}, for small $\theta$, the scattering amplitude will be dominated by the exchange of soft particles. The force mediated by the soft particle exchange with mass $M$ is expected to decay as $\exp[-Mb]$ for large impact parameter $Mb\gg1$. 
On the other hand, scattering with fixed $b$ will grow as $\exp[Cs^\alpha]$ when the scattering energy increases, because the number of excited intermediate multi-particle states can increase as $\exp[Cs^\alpha]$ at most, which is responsible for an exponential growth of the Lehmann-K{{\"{a}}ll{\'{e}}n spectral density. With taking into account $l\sim q_sb$, the above discussion implies the partial wave amplitudes may behave as
\begin{align}
\left|f_l(s)\right|\sim e^{Cs^\alpha}e^{-\frac{M}{q_s}l}\,,
\end{align}
for sufficiently large $l$. Then, using $q_s\sim\sqrt s$, exponential suppression of $f_l(s)$ for $l\gtrsim \tilde L(s)\coloneqq M^{-1}Cs^{\alpha+\frac{1}{2}}$ is expected. Thus, 
\begin{align}
F^{(n)}(s)\sim\sum_{l=0}^{\tilde L(s)}(2l+1)f_l(s)\left(\frac{l}{q_s}\right)^{2n}\,.
\end{align}
Unitarity in the form of $SS^\dagger=1$ which leads to the condition $\left|f_l(s)\right|\leq1$ further constrains the high energy behavior of the scattering amplitude:
\begin{align}
&\left|F^{(n)}(s)\right|\sim\sum_{l=0}^{\tilde L(s)}(2l+1)\left(\frac{l}{q_s}\right)^{2n}\sim \left({\tilde L(s)}\right)^{2n+2}s^{-n}=\left(M^{-1}C\right)^{2n+2}s^{1+(2n+2)\alpha}\,.
\end{align}
This coincides with the obtained bounds \eqref{realsbound1} assuming the holomorphy of $\tilde F(s,Z)$ inside the ellipse $\mathcal C_0$. This result indicates the holomorphy of $F(s,t)$ inside the ellipse $\mathcal C_0$ is intimately related to the short-range nature of the force. 
\section{Positivity bounds}\label{secposibound1}
In this section, we derive the positivity bounds for $\alpha>0$ case. 
In sec.~\ref{subsecposi1}, we review the derivation of positivity bounds for $\alpha=0$ case based on \cite{deRham:2017avq, deRham:2017imi}. In sec.~\ref{subsecposi2}, we derive the positivity bounds for $\alpha>0$ case. It turns out that it seems impossible to obtain any positivity bounds for $\alpha\geq1$ case, because we fail to obtain the dispersion relation with a finite number of subtractions. In sec.~\ref{subsecposi3}, we obtain the criterion which allows us to obtain the lower bound on $\alpha$ through the violation of positivity bounds. We also briefly comment on the application of our results to massive Galileon model and its massless limit.
\subsection{Positivity bounds for tempered localizable field}\label{subsecposi1}
Using Cauchy's integral formula, $F(s,t)$ can be expressed as
\begin{align}
F(s,t)=\left(s-2m^2+\frac{t}{2}\right)^{2}\oint_{\mathcal C}\frac{\mathrm{d}s'}{2\pi i}\frac{F(s',t)}{\left(s'-s\right)\left(s'-2m^2+\frac{t}{2}\right)^{2}}\,,
\end{align}
where $F(s,t)$ is holomorphic in $s$ inside the counterclockwise contour $\mathcal{C}$. Then, by deforming the contour and fully making use of the $s\leftrightarrow u$ crossing symmetry, one obtains
\begin{align}
F(s,t)&=\left[-\frac{{\rm Res}_{s=m^2}F(s,t)}{m^2-s}+\frac{{\rm Res}_{u=m^2}F(s,t)}{m^2-u}\right]+\sum_{k=0}^{1}a_{k}(t)s^{k}\no\\
&+\frac{2\left(\bar s+\frac{\bar t}{2}\right)^{2}}{\pi}\int^\infty_{4m^2}\mathrm{d}\mu\frac{\im\, F\left(\mu+i\epsilon,t\right)}{\left(\bar\mu+\frac{\bar t}{2}\right)\left[\left(\bar\mu+\frac{\bar t}{2}\right)^2-\left(\bar s+\frac{\bar t}{2}\right)^2\right]}\,,\label{disp1}
\end{align}
where $\bar X\coloneqq X-\bigl(4m^2/3\bigr)$. Here, we used the fact that $\bigl|s^{-2}\cdot F(s,t)\bigr|\to0$ as $|s|\to\infty$ for $0\leq t< 4m^2$, $t\neq m^2$ when $\alpha=0$ \cite{Jin:1964zza}. We also used the Schwarz reflection principle. We did not write down the explicit expressions of coefficients $a_{k}(t)$ because they are irrelevant for positivity bounds. 
Then, the pole-subtracted amplitude $B(s,t)$ which is defined by
\begin{align}
B(s,t)\coloneqq F(s,t)-\left[\frac{-{\rm Res}_{s=m^2}F(s,t)}{m^2-s}+\frac{{\rm Res}_{u=m^2}F(s,t)}{m^2-u}+\frac{{\rm Res}_{t=m^2}F(s,t)}{m^2-t}\right]\,,
\end{align}
can be expressed as 
\begin{align}
B(s,t)&=\frac{{\rm Res}_{t=m^2}F(s,t)}{m^2-t}+\sum_{k=0}^{1}a_{k}(t)s^{k}\no\\
&+\frac{2\left(\bar s+\frac{\bar t}{2}\right)^{2}}{\pi}\int^\infty_{4m^2}\mathrm{d}\mu\frac{\im\, F\left(\mu+i\epsilon,t\right)}{\left(\bar \mu+\frac{\bar t}{2}\right)\left[\left(\bar \mu+\frac{\bar t}{2}\right)^2-\left(\bar s+\frac{\bar t}{2}\right)^2\right]}\,,\label{posi0}
\end{align}
for $0\leq t< 4m^2$, $t\neq m^2$. Thus, by making use of the fact that ${\rm Res}_{t=m^2}F(s,t)$ is independent of $s$ in scalar theories, one can derive the following expression from eq.~\eqref{posi0}: 
\begin{align}
&B^{(2N,M)}(t)=\sum_{k=0}^{M}\frac{(-1)^k}{k!2^k}I^{(2N+k, M-k)}(t)\,,\label{posi1a}
\end{align}
for $0\leq t< 4m^2$, $t\neq m^2$, with $N\geq1$ and $M\geq0$. Here, $B^{(2N,M)}(t)$ and $I^{(q,p)}(t)$ are defined by 
\begin{align}
&B^{(2N,M)}(t)\coloneqq\left.\frac{\der^{2N}_v\der^M_t}{M!}\tilde B(v,t)\right|_{v=0}\,,\quad \tilde B(v,t)\coloneqq \left.B(s,t)\right|_{s=v+2m^2-(t/2)}\,,\\
&I^{(q,p)}(t)\coloneqq\frac{q!2}{p!\pi}\int^\infty_{4m^2}\mathrm{d}\mu\frac{\der^{p}_t\im\, F\left(\mu+i\epsilon,t\right)}{\left(\bar\mu+\frac{\bar t}{2}\right)^{q+1}}>0\,.\label{posi5b}
\end{align}
In \eqref{posi5b}, we used $\der^n_t\im\, F(s+i\epsilon,t)>0$ for $n\geq0$, when $0\leq t<4m^2$, $t\neq m^2$, and $s\geq 4m^2$ are simultaneously satisfied \cite{deRham:2017imi}. Note that the right-hand-side of eq.~\eqref{posi5b} is finite thanks to the bounds \eqref{frois1}. $B^{(2N,M)}(t)$ is however not ensure to be positive for $M\geq1$ because $(-1)^k$ factor is included in the right-hand side of \eqref{posi1a}. Then, from eqs.~\eqref{posi1a} and \eqref{posi5b} together with
\begin{align}
I^{(q,p)}(t)<\frac{q}{\mathcal M^2}I^{(q-1,p)}(t)\,,\label{Irel1}
\end{align}
where $\mathcal{M}^2\coloneqq\bigl(\bar \mu+(\bar t/2)\bigr)_{\mu=\mu_{\rm min}}$, $\mu_{\rm min}$ is the lower bound value of $\mu$ of the integration over $\mu$ on the right-hand side of eq.~\eqref{posi5b}, it has been shown that the following infinite number of inequalities which are recursively defined must be satisfied \cite{deRham:2017avq, deRham:2017imi}:
\begin{subequations}
\label{posi2}
\begin{align}
&Y^{(2N,0)}(t)\coloneqq B^{(2N,0)}(t)\,,\label{posi2a}\\
&Y^{(2N,M)}(t)\coloneqq\sum_{r=0}^{M/2}c_rB^{(2N+2r,M-2r)}(t)+\frac{1}{\mathcal{M}^2}\sum^{(M-1)/2}_{{\rm even}\,k=0}\left(2(N+k)+1\right)\beta_kY^{(2(N+k),M-2k-1)}(t)>0\,,\label{posi2b}
\end{align}
\end{subequations}
for $N\geq1, M\geq0$ and $0\leq t<4m^2\,,\,t\neq m^2$. 
Coefficients $c_k$ and $\beta_k$ are defined by
\begin{align}
c_k\coloneqq\frac{E_{2k}}{(2k)!2^{2k}}\,,\quad \beta_k\coloneqq\frac{(-1)^k\left(2^{2k+3}-2\right)B_{2k+2}}{(2k+2)!}\,,
\end{align}
where $E_{2k}$ and $B_{2k+2}$ denote the Euler numbers and Bernoulli numbers, respectively. These bounds are also extended to the scattering of the particles with spin \cite{deRham:2017zjm, deRham:2018qqo} and are applied to various models ({\it e.g.}, \cite{Bonifacio:2016wcb, deRham:2017imi, deRham:2017xox, Bellazzini:2017fep}). An infinite number of inequalities \eqref{posi2} with $N\geq1$, $M\geq0$ are obtained, and they are called positivity bounds. Thanks to the condition $v=0$ and $0\leq t<4m^2$, the left-hand side of \eqref{posi2} can be evaluated within the validity of the LEEFT, meaning that \eqref{posi2} gives the non-trivial bounds on LEEFT.

In fact, one can improve the above positivity bounds eq.~\eqref{posi2} when perturbation theory based on LEEFT is valid for $s<\Lambda_{\rm th}^2$ with $m^2\ll\Lambda_{\rm th}^2$, by evaluating $\im F(\mu+i\epsilon)$ with $4m^2<\mu<\Lambda_{\rm th}^2$ which expresses the contributions from loops of light particle $\phi$ with external momenta satisfying $4m^2<s<\Lambda_{\rm th}^2$:
\begin{subequations}
\label{improved1}
\begin{align}
&B^{(2N,M)}_{\Lambda_{\rm th}}(t)\coloneqq B^{(2N,M)}(t)-\sum_{k=0}^M\frac{(-1)^k}{k!2^k}\frac{(2N+k)!2}{(M-k)!\pi}\int^{\Lambda_{\rm th}^2}_{4m^2}\mathrm{d}\mu\frac{\der^{M-k}_t\im\, F\left(\mu+i\epsilon,t\right)}{\left(\bar\mu+\frac{\bar t}{2}\right)^{2N+k+1}}\no\\
&\qquad\qquad=\sum_{k=0}^M\frac{(-1)^k}{k!2^k}I^{(2N+k,M-k)}_{\Lambda_{\rm th}}(t)\,,\label{improved1a}\\
&I^{(q,p)}_{\Lambda_{\rm th}}(t)\coloneqq\frac{q!2}{p!\pi}\int^\infty_{\Lambda_{\rm th}^2}\mathrm{d}\mu\frac{\der^{p}_t\im\, F\left(\mu+i\epsilon,t\right)}{\left(\bar\mu+\frac{\bar t}{2}\right)^{q+1}}>0\,.
\end{align}
\end{subequations}
From the definition of $\Lambda_{\rm th}$, one can evaluate $B^{(2N,M)}_{\Lambda_{\rm th}}(t)$ perturbatively by using LEEFT . Then, by making use of inequalities
\begin{align}
I^{(q,p)}_{\Lambda_{\rm th}}(t)<\frac{q}{\Lambda_{\rm th}^2+(t/2)-2m^2}I^{(q-1,p)}_{\Lambda_{\rm th}}(t)
\,,\label{Irel2}
\end{align}
which is just an analogue of \eqref{Irel1}, one obtains the improved positivity bounds:
\begin{subequations}
\label{posi3}
\begin{align}
&Y_{\Lambda_{\rm th}}^{(2N,0)}(t)\coloneqq B_{\Lambda_{\rm th}}^{(2N,0)}(t)\,,\label{posi3a}\\
&Y_{\Lambda_{\rm th}}^{(2N,M)}(t)\coloneqq\sum_{r=0}^{M/2}c_rB_{\Lambda_{\rm th}}^{(2N+2r,M-2r)}(t)\no\\
&\qquad\qquad+\frac{1}{\Lambda_{\rm th}^2+(t/2)-2m^2}\sum^{(M-1)/2}_{{\rm even}\,k=0}\left(2(N+k)+1\right)\beta_kY_{\Lambda_{\rm th}}^{(2(N+k),M-2k-1)}(t)>0\,.\label{posi3b}
\end{align}
\end{subequations}

\subsection{Positivity bounds without temperedness assumption}
In this subsection, we derive the positivity bounds without temperdness assumption, namely, for theories with $\alpha>0$. 
\label{subsecposi2}
\subsubsection*{\textbullet\,\,\,\, $0<\alpha<1$ case}
Using Cauchy's integral formula, $F(s,t)$ can be expressed as
\begin{align}
F(s,t)=\left(s-2m^2+\frac{t}{2}\right)^{2P}\oint_{\mathcal C}\frac{\mathrm{d}s'}{2\pi i}\frac{F(s',t)}{\left(s'-s\right)\left(s'-2m^2+\frac{t}{2}\right)^{2P}}\,,
\end{align}
for any $P\in\mathbb N$. Here $F(s,t)$ is holomorphic in $s$ inside the counterclockwise contour $\mathcal{C}$. Then, combined with an analyticity assumption B2, one obtains the following expression by deforming the contour and fully making use of the $s\leftrightarrow u$ crossing symmetry:
\begin{align}
F(s,t)&=\left[-\frac{{\rm Res}_{s=m^2}F(s,t)}{m^2-s}+\frac{{\rm Res}_{u=m^2}F(s,t)}{m^2-u}\right]+\sum_{k=0}^{2P-1}b_{k}(t)s^{k}\no\\
&+\frac{2\left(\bar s+\frac{\bar t}{2}\right)^{2P}}{\pi}\int^R_{4m^2}\mathrm{d}\mu\frac{\im\, F\left(\mu+i\epsilon,t\right)}{\left(\bar\mu+\frac{\bar t}{2}\right)^{2P-1}\left[\left(\bar\mu+\frac{\bar t}{2}\right)^2-\left(\bar s+\frac{\bar t}{2}\right)^2\right]}\no\\
&+\frac{\left(\bar s+\frac{\bar t}{2}\right)^{2P}}{\pi}\int^{R+4{m^2}-t}_R\mathrm{d}\mu\frac{\im\, F\left(\mu+i\epsilon,t\right)}{\left(\bar\mu+\frac{\bar t}{2}\right)^{2P}\left[\left(\bar\mu+\frac{\bar t}{2}\right)+\left(\bar s+\frac{\bar t}{2}\right)\right]}\no\\
&+\left(\bar s+\frac{\bar t}{2}\right)^{2P}\int_{{\mathcal C}^\pm_{R}}\frac{\mathrm{d}s'}{2\pi i}\frac{F(s',t)}{\left(s'-s\right)\left(\bar{s'}+\frac{\bar t}{2}\right)^{2P}}\,,\label{disp2}
\end{align}
where $\mathcal C^{\pm}_R$ denotes the semi-circle with radius $R$: $\mathcal C^+_R\coloneqq \bigl\{s:s=Re^{i\theta}+i\epsilon,\,0\leq\theta\leq\pi\bigr\}$ and $\mathcal C^-_R\coloneqq \bigl\{s:s=Re^{i\theta}-i\epsilon,\,\pi\leq\theta\leq2\pi\bigr\}$. We did not write down the explicit expressions of coefficients $b_{k}(t)$ because they are irrelevant for positivity bounds. We will take the limit $R\to\infty$ later. For any $s\in\mathbb C$ in the complex-s plane, except for poles and branch cuts on the real-$s$ axis, $F(s,t)$ is analytic in $t$ at $t=0$ from the analyticity assumption. Thus, eq.~\eqref{disp2} gives rise to
\begin{align}
\left.\frac{\der^n_t}{n!}\tilde F(v,t)\right|_{t=0}&=\frac{\der^n_t}{n!}\left[\left(-\frac{{\rm Res}_{s=m^2}F(s,t)}{m^2-s}+\frac{{\rm Res}_{u=m^2}F(s,t)}{m^2-u}+\sum_{k=0}^{2P-1}b_{k}(t)s^{k}\right)_{s=v+2m^2-(t/2)}\right]_{t=0}\no\\
&+\frac{2v^{2P}}{n!\pi}\int^R_{4m^2}\mathrm{d}\mu\,\der^n_t\left\{\frac{\im\, F\left(\mu+i\epsilon,t\right)}{\left(\bar\mu+\frac{\bar t}{2}\right)^{2P-1}\left[\left(\bar\mu+\frac{\bar t}{2}\right)^2-v^2\right]}\right\}_{t=0}\no\\
&+\frac{v^{2P}}{n!\pi}\der^n_t\left[\int^{R+4{m^2}-t}_R\mathrm{d}\mu\,\frac{\im\, F\left(\mu+i\epsilon,t\right)}{\left(\bar\mu+\frac{\bar t}{2}\right)^{2P}\left(\bar\mu+\frac{\bar t}{2}+v\right)}\right]_{t=0}\no\\
&+\frac{v^{2P}}{n!}\int_{{\mathcal C}^\pm_{R}}\frac{\mathrm{d}s'}{2\pi i}\,\der^n_t\left[\frac{F(s',t)}{\left(\bar{s'}+\frac{\bar t}{2}-v\right)\left(\bar{s'}+\frac{\bar t}{2}\right)^{2P}}\right]_{t=0}\,,\label{disp3}
\end{align}
where
\begin{align}
\tilde F(v,t)\coloneqq \left.F(s,t)\right|_{s=v+2m^2-(t/2)}\,.
\end{align}
Thus,
\begin{align}
&B^{(2N,M)}(0)=\left[\der^{2N}_v\left(\left.\frac{\der^M_t}{M!}\tilde B(v,t)\right|_{t=0}\right)\right]_{v=0}\no\\
&=\frac{(2N)!2}{M!\pi}\int^R_{4m^2}\mathrm{d}\mu\,\der^M_t\left[\frac{\im\, F\left(\mu+i\epsilon,t\right)}{\left(\bar\mu+\frac{\bar t}{2}\right)^{2N+1}}\right]_{t=0}+\left.\der_v^{2N}\left[\left.\frac{\der_t^M}{M!}\left(\frac{\left.{\rm Res}_{t=m^2}F(s,t)\right|_{s=v+2m^2-(t/2)}}{m^2-t-i\epsilon}\right)\right|_{t=0}\right]\right|_{v=0}\no\\
&+\frac{(2N)!}{M!\pi}\der^M_t\left[\int^{R+4{m^2}-t}_R\mathrm{d}\mu\,\frac{\im\, F\left(\mu+i\epsilon,t\right)}{\left(\bar\mu+\frac{\bar t}{2}\right)^{2N+1}}\right]_{t=0}+\frac{(2N)!}{M!}\int_{{\mathcal C}^\pm_{R}}\frac{\mathrm{d}s'}{2\pi i}\,\der^M_t\left[\frac{F(s',t)}{\left(\bar{s'}+\frac{\bar t}{2}\right)^{2N+1}}\right]_{t=0}\,,\label{subtdisp2}
\end{align}
for any $N\in\mathbb N$ and $M\geq0$. Note that when ${\rm Res}_{t=m^2}F(s,t)$ is independent of $s$, the second term in the second line of eq.~\eqref{subtdisp2} vanishes. In scalar theories, ${\rm Res}_{t=m^2}F(s,t)$ is independent of $s$, and hence one can drop the contribution from $t$-channel pole in \eqref{subtdisp2}. Now, let us consider what happens if we take the limit $R\to\infty$ in eq.~\eqref{subtdisp2}. From \eqref{realsbound1}, $s\leftrightarrow u$ crossing symmetry, and the Schwarz reflection principle, it turns out that $\bigl|F^{(n)}(s\pm\epsilon)\bigr|<|s|^{1+(2n+2)\alpha}$ along the real axis. If this polynomial boundedness property also holds on the complex $s$-plane, the terms in the final line of eq.~\eqref{subtdisp2} vanish after taking the limit $R\to\infty$. In order to show the boundedness property of $\bigl|F^{(n)}(s)\bigr|$ in the complex $s$-plane, the Phragm{\'{e}}n-Lindel{\"{o}}f theorem plays a crucial role. Combined with the analyticity assumption B2, the theorem ensures that the polynomial boundedness \eqref{realsbound1} on the real axis also holds for modulus of $s$ in the upper-half plane if $\bigl|F^{(n)}(s)\bigr|$ can grow no faster than $\exp[\sigma|s|^\gamma]$ with $\gamma<1$ as $|s|\to\infty$ in the upper-half plane. Combined with the boundedness assumption C, this means that for $\alpha<1$ case, $\bigl|F^{(n)}(s)\bigr|$ is bounded by $|s|^{1+(2n+2)\alpha}$ in the upper-half plane.\footnote{This polynomial boundedness on the scattering amplitudes in the complex $s$-plane is consistent with the result obtained in \cite{Epstein:1969bg}.} This boundedness in the upper-half plane also ensures the boundedness in the lower-half plane by making use of the $s\leftrightarrow u$ crossing symmetry. Thus, from eqs.~\eqref{realsbound1} and \eqref{subtdisp2}, one can obtain the following expression by taking the limit $R\to\infty$, {\it only} for $2N>1+(2+2M)\alpha$:
\begin{align}
B^{(2N,M)}(0)=\frac{(2N)!2}{M!\pi}\int^\infty_{4m^2}\mathrm{d}\mu\,\der^M_t\left[\frac{\im\, F\left(\mu+i\epsilon,t\right)}{\left(\bar\mu+\frac{\bar t}{2}\right)^{2N+1}}\right]_{t=0}=\sum_{k=0}^{M}\frac{(-1)^k}{k!2^k}I^{(2N+k,M-k)}(0)\,.\label{subtdisp3}
\end{align}
This means that for $0<\alpha<1$ case, one can relate $B^{(2N,M)}(t)$ to the imaginary part of the scattering amplitude as in eq.~\eqref{posi1a} only when both $2N>1+(2M+2)\alpha$ and $t=0$ are satisfied. Therefore, for $0<\alpha<1$ case, positivity bounds \eqref{posi2} with $2N>1+(2M+2)\alpha$ and $t=0$ must be still satisfied, while other bounds which must be satisfied in $\alpha=0$ case could be violated. Indeed, eqs.~\eqref{realsbound1}, \eqref{posi1a}, \eqref{posi5b}, and \eqref{posi2} imply that for $\alpha>0$ case, $I^{(2N,M)}(t)$, $B^{(2N,M)}(t)$, and $Y^{(2N,M)}(t)$ are ensured to be finite only when both $2N>1+(2M+2)\alpha$ and $t=0$ are satisfied. Note that one can also obtain the improved positivity bounds by introducing $\Lambda_{\rm th}$ which specifies the regime of the validity of the perturbative calculation of LEEFT, and evaluating the contributions from the loops of light particle as is done for $\alpha=0$ case in sec.~\ref{subsecposi1}.

The result obtained here is remarkable: {\it there are still infinite number of inequalities which must be satisfied by LEEFT scattering amplitude, even when UV completions are non-local.} Existence of such bounds implies the existence of IR obstructions to unitary, analytic, and Lorentz invariant UV completions. In order to demonstrate the importance of this result, let us consider the forward-limit positivity bounds $Y^{(2N,0)}(0)>0$ with $N\geq1$, for example. Leading-order forward-limit positivity bound $Y^{(2,0)}(0)>0$ can be an IR obstruction for LEEFTs to admit local, analytic, unitary, and Lorentz invariant UV completions, even if LEEFTs are apparently consistent with locality and Lorentz invariance. This statement has been already obtained in \cite{Adams:2006sv}.
In the literature, sub-leading order forward-limit positivity bounds $Y^{(2N,0)}(0)>0$ with $N\geq2$ are also regarded as IR obstructions to {\it local}, analytic, unitary, and Lorentz invariant UV completions. However, our results suggest that sub-leading order forward-limit positivity bounds can be IR obstructions to analytic, unitary, and Lorentz invariant but possibly {\it non-local} UV completions. 
\begin{figure}[tbp]
 \centering
  \includegraphics[width=.7\textwidth,trim=120 150 130 90,clip]{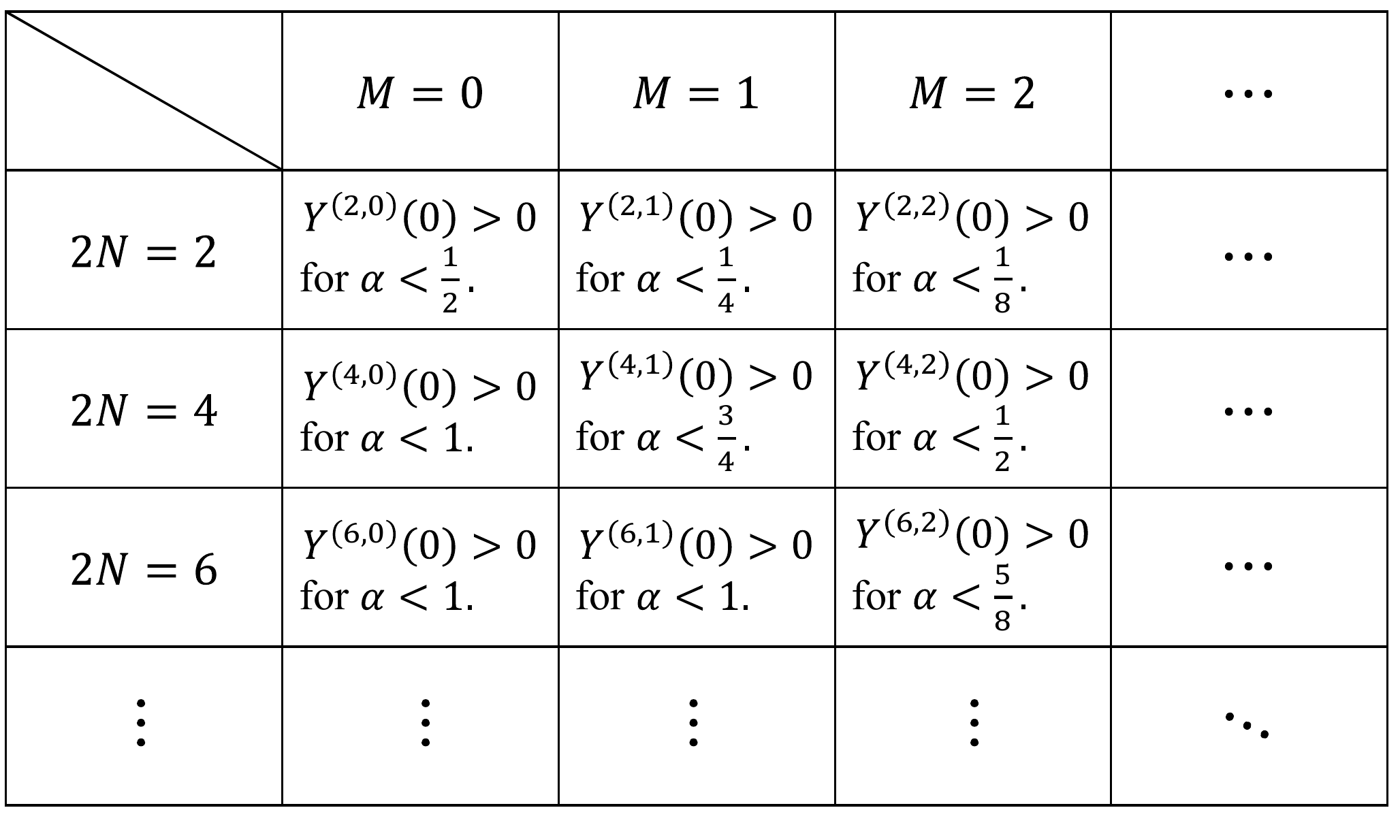}
 \caption{This figure shows which inequalities must be satisfied for generic $\alpha>0$. For $\alpha>0$ case, all the bounds could be violated when $t>0$. When $t=0$, some of the bounds must still be satisfied for $0<\alpha<1$ case: for fixed $M$ and $\alpha$ with $0<\alpha<1$, inequalities with $2N>1+(2+2M)\alpha$ must be satisfied. For fixed $N$ and $\alpha$ with $0<\alpha<1$, inequalities with $M\geq (2\alpha)^{-1}(2N-1)-1$ could be violated even when $t=0$. When $\alpha\geq1$, all the inequalities could be violated. Of course, one can also derive the improved positivity bounds by introducing the parameter $\Lambda_{\rm th}$ and evaluating the loop corrections from the light particle.}
 \label{figboundlist1} 
\end{figure}
 
\subsubsection*{\textbullet\,\,\,\,$\alpha\geq1$ case}\label{subsubsecbound3}
As is discussed just before obtaining eq.~\eqref{subtdisp3}, one cannot make use of Phragm{\'{e}}n-Lindel{\"{o}}f theorem for $\alpha\geq1$ case, because $\bigl|F^{(n)}(s)\bigr|$ can grow as fast as $\exp\bigl[C|s|\bigr]$ in this case. This means that it would be impossible to show the polynomial boundedness of $\bigl|F^{(n)}(s)\bigr|$ in the complex-$s$ plane even if $\bigl|F^{(n)}(s)\bigr|$ is polynomially bounded on the real axis. This means that it will be impossible to obtain dispersion relation with a finite number of subtractions for $\alpha\geq1$ case, and hence all the positivity bounds can be violated.\\

Let us summarize the results obtained in this subsection \ref{subsecposi2}. Among an infinite number of positivity bounds \eqref{posi2} which must be satisfied in $\alpha=0$ case, all the bounds with $t>0$ could be violated for any value of $(2N,M)$ in $\alpha>0$ case. On the other hand, some of positivity bounds \eqref{posi2} with $t=0$ must be still satisfied for $0<\alpha<1$. In fig.~\ref{figboundlist1}, we list up which inequalities must be satisfied for generic $\alpha$. For fixed $M$ and $\alpha$ with $0<\alpha<1$, inequalities with $2N>1+(2+2M)\alpha$ must be satisfied. For fixed $N$ and $\alpha$ with $0<\alpha<1$, inequalities with $M\geq (2\alpha)^{-1}(2N-1)-1$ could be violated even when $t=0$.  When $\alpha\geq1$, all the inequalities could be violated. As is emphasized above, there are still infinite number of positivity bounds for $\alpha<1$, such as sub-leading order forward-limit positivity bounds. This implies the existence of IR obstructions to Lorentz invariant UV completions: {\it our results open the new possibility to falsify unitary, analytic, and Lorentz invariant UV completions via the violation of positivity bounds, even if LEEFT is apparently Lorentz invariant.}

\subsection{Lower bound on the $\alpha$ parameter}\label{subsecposi3}
If we maintain the assumptions A-C which are given in sec.~\ref{assumption1}, the results obtained in sec.~\ref{subsecposi1}-\ref{subsecposi2} allow us to put a lower bound on $\alpha$ in the following manner:
\begin{itemize}
\item If at least one of the positivity bounds \eqref{posi2}  is violated for $t>0$, then $\alpha>0$.
\item If the positivity bounds \eqref{posi2} with $t=0$ are violated for $\forall(2N,M)\in\{(2N_i,M_i)\}_{i\in \mathcal{D}}$, where $\mathcal D$ specifies the set of variables $(2N,M)$, then 
\begin{align}
\alpha\geq{\rm Min}\left[\frac{2N_*-1}{2M_*+2}\,, 1\right]\,,\quad \frac{2N_*-1}{2M_*+2}\coloneqq{\rm Max}\left[\left\{\frac{2N_i-1}{2M_i+2}\right\}_{i\in\mathcal{D}}\right]\,.\label{criterion1}
\end{align}
\end{itemize}
This result shows that one can obtain the information of the $\alpha$ parameter only from the low energy data which would be accessible from observations, by investigating which inequalities are violated among an infinite number of inequalities \eqref{posi2}. 

\subsubsection*{Application to massive Galileon: massless limit}\label{secgal}
Let us briefly comment on the application of our results to massive Galileon models. Positivity bounds are applied to massive Galileon models in \cite{deRham:2017imi}. Their results suggest that it is impossible to take the massless limit with satisfying the leading-order forward limit positivity bound. From eq.~\eqref{criterion1}, this means that the UV completion of Galileon models which are obtained by taking the massless limit of massive Galileon models cannot be strictly localizable theories: if one try to maintain unitarity, analyticity, and Lorentz invariance, one must give up strict locality. This is consistent with the conjecture proposed in \cite{Keltner:2015xda} that Galileons may fall into non-localizable theory.
\section{Conclusion and Discussions}\label{concl}
In this study, we have derived the posivity bounds on LEEFT which admit an analytic, unitary, and Lorentz invariant UV completions with a mass gap. 
Under several reasonable assumptions on the S-matrix, we find that an infinite number of subtractions will be required for $t>0$, unless $\alpha=0$. This means that the beyond forward limit positivity bounds obtained in the literature can be violated even in local QFT because standard local QFT can have $0\leq \alpha<\frac{1}{2}$. Beyond forward limit positivity bounds are obtained only when $\alpha=0$ case, namely, momentum space Wightman functions are polynomially bounded. 
On the other hand, we have shown that $\der_t^n F(s,t)\bigr|_{t=0}$ are polynomially bounded in the complex $s$-plane when $\alpha<1$, leading to the dispersion relation with finite number of subtractions. As a result, we obtained an infinite number of positivity bounds for $\alpha<1$ theories: obtained bounds for generic $\alpha$ are summarized in fig.~\ref{figboundlist1}. Because $\alpha\geq\frac{1}{2}$ theories are essentially non-local, our results suggest the existence of IR obstructions to analytic, unitary, and Lorentz invariant UV completions. 
So far, it was impossible to conclude that the UV completion must violate at least one of the analyticity, unitarity, or Lorentz invariance, by testing the violation of positivity bounds. This is because it was not known whether positivity bounds could be obtained when the locality assumption is removed. {\it Our results open the very exciting windows to test the analyticity, unitarity, or Lorentz invariance of UV completion only from IR data, even if Lorentz invariance is maintained at low energy.} 

As is well known, however, one of the most annoying issue is the necessity of the introduction of a mass gap in the scattering amplitude under consideration to derive positivity bounds. The existence of a mass gap is important for deriving the high energy behavior of the amplitude and the dispersion relation. Both of them are crucially important to derive positivity bounds. Because massless graviton will be coupled to all sectors, this issue must be considered seriously. 
It should be clarified under which situation one can correctly capture the properties of the amplitude by introducing mass terms in the theory as an IR regulator. 
We will study these aspects in our future work.


\acknowledgments
J. T. would like to thank Andrew J. Tolley and Claudia de Rham for useful discussions and valuable comments, as well as their hospitality during his visits to Imperial College London. The author also would like to thank Takahiro Tanaka for useful discussions.


 \bibliography{positivity.bib}

\end{document}